\documentclass[twoside]{article}

\usepackage{PRIMEarxiv}

\usepackage[utf8]{inputenc} % allow utf-8 input
\usepackage[T1]{fontenc}    % use 8-bit T1 fonts
\usepackage{hyperref}       % hyperlinks
\usepackage{url}            % simple URL typesetting
\usepackage{booktabs}       % professional-quality tables
\usepackage{amsfonts}       % blackboard math symbols
\usepackage{nicefrac}       % compact symbols for 1/2, etc.
\usepackage{microtype}      % microtypography
\usepackage{lipsum}
\usepackage{fancyhdr}       % header
\usepackage{graphicx}       % graphics
\graphicspath{{media/}}     % organize your images and other figures under media/ folder
\usepackage{subcaption}

\title{``I'm Not Confident in Debiasing AI Systems Since I Know Too Little'': Teaching AI Creators About Gender Bias Through Hands-on Tutorials}

\author{
  Kyrie Zhixuan Zhou\textsuperscript{1},
  Jiaxun Cao\textsuperscript{2},
  Xiaowen Yuan\textsuperscript{3},
  Daniel E. Weissglass\textsuperscript{2},
  \\ \textbf{Zachary Kilhoffer\textsuperscript{1},}
  \textbf{Madelyn Rose Sanfilippo\textsuperscript{1},}
  \textbf{Xin Tong\textsuperscript{2}}\\
  \textsuperscript{1}University of Illinois at Urbana-Champaign\\
  \{zz78, dzk2, madelyns\}@illinois.edu\\
  \textsuperscript{2}Duke Kunshan University\\
  \{jessie.cao, xin.tong\}@dukekunshan.edu.cn, daniel.weissglass@duke.edu \\
  \textsuperscript{3}University of California, Berkeley\\
  xiaowen\_yuan@berkeley.edu
}

\begin{document}
\maketitle

\begin{abstract}
  Gender bias is rampant in AI systems, causing bad user experience, injustices, and mental harm to women. School curricula fail to educate AI creators on this topic, leaving them unprepared to mitigate gender bias in AI. In this paper, we designed hands-on tutorials to raise AI creators' \textit{awareness} of gender bias in AI and enhance their \textit{knowledge} of sources of gender bias and debiasing techniques. The tutorials were evaluated with 18 AI creators, including AI researchers, AI industrial practitioners (i.e., developers and product managers), and students who had learned AI. Their improved awareness and knowledge demonstrated the effectiveness of our tutorials, which have the potential to complement the insufficient AI gender bias education in CS/AI courses. Based on the findings, we synthesize design implications and a rubric to guide future research, education, and design efforts. 
\end{abstract}

\keywords{AI \and Gender bias \and Education \and Tutorial}

\section{Introduction}

Gender bias refers to a person receiving different treatment based on the person's real or perceived gender identity \cite{jacobson1992gender}. Gender bias covers four main categories \cite{dev2021measures}, including \textit{stereotyping}, i.e., overgeneralized beliefs about the personal attributes of an individual as determined by their demographic group membership \cite{ellemers2018gender}, \textit{disparagement}, i.e., the notion that certain groups are less valuable than others and less deserving of respect \cite{abrams2015gender}, \textit{dehumanization}, i.e., casting disfavored groups as ``others'' and aiming to erase signs of shared humanity \cite{tipler2019dehumanizing}, and \textit{erasure}, i.e., the lack of adequate representation of members of a particular social group \cite{mibenge2013sex}.

Gender bias broadly exists in AI tools for recruitment \cite{dastin2022amazon},  recommendation \cite{gender:bias:5}, and many other scenarios that people regularly encounter \cite{broussard2023more}. As a result, procedural and substantive injustices \cite{endo2018technological} frequently arise against women. For example: women receive fewer job posting advertisements on social media \cite{lambrecht2019algorithmic}; medical AI is less effective for women \cite{cirillo2020sex}; and Google Translate exhibits a strong tendency toward male defaults, particularly for fields typically associated with unbalanced gender distribution or stereotypes such as STEM jobs \cite{prates2020assessing}. 

Gender bias in AI thus remains a pressing issue, and AI creators are the ones who are responsible for building bias-free AI systems. Yet, AI creators often lack knowledge of preventive measures to avoid bias in AI applications \cite{gender:bias:14}. Male AI creators, who account for a significant portion of this workforce, have less awareness of gender bias than female creators \cite{leavy2018gender, medel2017eliminating}.
Equipping AI creators with both awareness and knowledge of AI gender bias is timely and important. However, little research has been devoted to understanding how and how well AI gender bias is taught in CS/AI courses \cite{bias:education:8, medel2017eliminating}. To our knowledge, there is only one AI gender bias education tool devoted to teaching this topic to children and youth \cite{bias:education:1}, effectively raising their awareness of gender bias in AI. However, this tool is not suitable for AI creators who need more technical knowledge such as debiasing techniques. The best way to introduce the topic of AI gender bias to AI creators remains under-investigated. 

To bridge this research gap, we designed hands-on tutorials to teach AI gender bias to AI creators and equip them with awareness and practical knowledge. At the awareness level, we intended to help them better recognize gender bias in AI and motivate them to solve this issue. At the knowledge level, we intended to convey technical knowledge, e.g., how gender bias is introduced into AI (sources of gender bias) and how gender bias can be mitigated from AI (debiasing techniques). To engage the learners, we designed the tutorials in a hands-on manner and used real-world scenarios, i.e., AI-based recruitment, which helps organizations effectively source and screen candidates \cite{dastin2022amazon}, and AI-based autocomplete, which search engines use to complete searches that users start to type \cite{karapapa2015search}, to facilitate the learning activities \cite{bias:education:1}. To make the education more meaningful for real-world AI development, we embedded technical knowledge (e.g., debiasing techniques) and components (e.g., code, dataset) into the tutorials.

Eighteen AI creators evaluated our tutorials, including four AI researchers who focused on the technical aspects of AI, four AI/HCI researchers who focused on the user/society aspects of AI, three AI developers, two AI product managers, and five students who had learned AI. Through interviews and user experiments, we aspired to answer the following research questions (RQs):
\begin{itemize}
    \item \textbf{RQ1:} How is current AI gender bias education delivered and received by AI creators?
    \item \textbf{RQ2:} Are our tutorials effective in increasing AI creators' awareness and knowledge of AI gender bias?
    \item \textbf{RQ3:} How do AI creators perceive the usability of our tutorials?
\end{itemize}

In the pre-study interview, the participants commonly expressed their lived experience of gender bias when using or creating AI, and indicated insufficient education in CS/AI courses and no education tools on this topic. The lack of education made them unable to identify gender bias in AI and mitigate gender bias from AI. After completing our tutorials, they showed an improvement in terms of both awareness and knowledge of AI gender bias. Awareness-wise, they expressed a strengthened ability to identify gender bias in AI and an enhanced intent to address the bias issue. Knowledge-wise, they perceived a heightened level of technical knowledge, evidenced by a higher accuracy rate in knowledge question (KQ) surveys, and felt more confident in debiasing AI. The participants expressed a generally good user experience of our tutorials and also suggested areas for further improvement.

The contribution of our work is thus three-fold. Firstly, we design and evaluate first-of-its-kind hands-on tutorials to raise AI creators' awareness and knowledge of AI gender bias. Secondly, we uncover the lack of AI gender bias education and aim to attract more attention from the AI, HCI, and CS Education community. Thirdly, we deliver design implications and a rubric to shed light on future research, education, and design practices.
\section{Related Work}
Our design draws on two lines of research, i.e., AI and AI ethics education in general, and AI gender bias education in classrooms and through education tools/materials.

\subsection{AI and AI Ethics Education}

AI is reshaping the world in profound ways, automating work and enhancing productivity \cite{manyika2017future}. To leverage its benefits and mitigate its potential harms \cite{kazim2021high}, systematic inclusion of AI ethics into the curriculum is important to get people prepared to face and solve ethical concerns such as security, privacy, and fairness \cite{borenstein2021emerging}.

A lot of research is focused on AI and AI ethics education in school and classroom settings, touching on topics taught, teaching methods used, and education effect \cite{bias:education:6, bias:education:7, bias:education:8, quinn2021readying, katznelson2021need, wright2020ai}. Ethical principles such as bias and privacy are taught in AI ethics courses and technical AI courses in some U.S. universities, though how they are conveyed and received by the students is unclear \cite{bias:education:8}. Green reported on the experience of teaching ethical principles and letting students incorporate the principles into agents in an undergraduate-level AI ethics course \cite{bias:education:6}. In addition to CS courses, AI ethics education is also integrated into the curricula of such fields as medicine \cite{quinn2021readying, katznelson2021need}, business \cite{de2023educating}, and law \cite{wright2020ai}, given the wide applications of AI and the corresponding high stakes. Even high school and middle school students could learn AI ethics such as epistemic norms, privacy, and digital citizenship in both CS and non-CS courses through discussion, gamified activities, content creation, and so on \cite{bias:education:7,eguchi2021contextualizing}.

Education tools are also designed and evaluated to teach AI and AI ethics to novice users, even for children and youth \cite{AI:education:2, AI:education:6, AI:education:4, AI:education:5}. Re-Shape was an educational environment for people to reflect on data privacy through collecting, processing, and visualizing their physical movement data \cite{bias:education:5}. Even youth with no programming experience could incorporate machine learning (ML) classifiers to model their physical activity interactively \cite{AI:education:4}.

The abundance of topics taught in AI and AI ethics education, as well as methods leveraged (e.g., hands-on approach \cite{AI:education:2, AI:education:6, AI:education:4, AI:education:5}), informed the design of our tutorials. 

\subsection{AI Gender Bias Education}

Gender bias has been repeatedly reported in AI applications such as emotion recognition \cite{gender:bias:3}, search engines \cite{gender:bias:4}, recommendation \cite{gender:bias:5}, and robots \cite{gender:bias:6}. Gender bias raises many concerns, but chief among these is that such bias will result in significant substantive and procedural injustices -- unfairness in the distribution of outcomes and the application of rules, respectively \cite{endo2018technological}. While either substantive or procedural injustice is concerning independently, they often occur together. For instance, experiments on advertising algorithms on a major social media platform showed that a job posting made in an explicitly ``gender neutral'' way -- specifying that it should be shown to both men and women -- was nonetheless more likely to be shown to men than women \cite{ali2019discrimination}. 
Women received fewer job posting advertisements because advertisements directed to women on that platform were competitive, which the researchers believed to result from preferential targeting of women for commercial advertisements \cite{lambrecht2019algorithmic}. In this case, we find a substantive injustice that women are presented with fewer job opportunities and targeted more aggressively for consumer spending. This substantive injustice resulted from procedural injustice -- even when directed to be ``gender neutral'', advertising algorithms effectively make decisions about whether to show someone a job opportunity based on their gender. Similar effects may result in medical AI being less able to care for women \cite{cirillo2020sex}, encoding a ``male default'' in machine translation \cite{prates2020assessing}, and disparity in a wide range of other domains \cite{gender:bias:1}. Given the proliferation of algorithmic systems in the contemporary world, such injustices can impact all parts of society and have costs for all facets of the human experience.

AI creators are supposed to debias AI. However, awareness and knowledge of AI gender bias are lacking in AI creators \cite{gender:bias:14,leavy2018gender, medel2017eliminating}. In a web-based survey, 45\% of the AI developers perceived their lack of knowledge as a reason for bias in AI systems, highlighting the need to strengthen knowledge about bias in AI, including concrete examples of gender bias, where biases in AI could occur, and preventive measures to avoid biases in AI \cite{gender:bias:14}. Male AI creators, who account for a significant portion of the AI workforce, have less awareness of gender bias than female creators \cite{leavy2018gender, medel2017eliminating}. Such cases highlight the need to strengthen AI creators' awareness and knowledge of AI gender bias. 

Despite the urgent need, AI gender bias education is not sufficiently investigated and delivered. Some of the 51 courses in US universities Garrett et al. examined contained gender bias as a topic, but omissions existed \cite{bias:education:8}. For example, diversity in the AI workforce was not called out in the syllabi for fixing the gender imbalance in tech companies. Further, they did not investigate how and how well AI gender bias was taught \cite{bias:education:8}, e.g., what approaches were used to convey the knowledge, and how much the students' knowledge and awareness of gender bias improved. These questions are key to assessing how useful these courses are in equipping AI creators with awareness and knowledge of AI gender bias. 

To our knowledge, only one educational tool has been designed to teach AI gender bias -- Melsión et al. developed an education platform to raise children's awareness of gender bias in AI by visually explaining AI predictions \cite{bias:education:1}. For example, the AI would classify a person in the kitchen as a woman for the ``feminine'' features such as kitchen utensils in an image. Though effective in helping children recognize gender bias in AI predictions, this tool is not suitable for AI creators who need technical knowledge in terms of sources of gender bias and debiasing techniques and who need an environment (e.g., dataset, code) mimicking real-world development practices.

\subsection{Research Gaps} 

Overall, AI gender bias is undercovered compared to other topics in AI/AI ethics education such as cybersecurity and privacy \cite{bias:education:7}. How effectively CS/AI courses teach gender bias is also under-investigated \cite{bias:education:8}. Existing research and practice fail to deliver a comprehensive AI gender bias education, including raising people's awareness (e.g., understanding and identifying gender bias in AI, having a mindset of solving the gender bias issue in AI), and enhancing their technical knowledge (e.g., sources of gender bias, debiasing techniques), which are lacking in AI creators \cite{gender:bias:14,leavy2018gender, medel2017eliminating}. 
To bridge these gaps, we designed hands-on tutorials on AI gender bias and conducted user experiments with AI creators to assess the tools. We also aimed to probe the status of AI gender bias education in CS/AI courses in the interviews.
\section{Designing Hands-on Tutorials For AI Gender Bias Education}

The overarching goals of our hands-on tutorials were to raise AI creators' awareness of AI gender bias and increase their knowledge of sources of gender bias and debiasing techniques. For simplicity, we treated gender as a binary construct like many prior studies on AI gender bias \cite{gender:bias:7}. Below, we elaborate on our design principles, content design of the education, and the design of two hands-on tutorials.

\subsection{Design Principles}
We synthesized several design principles to inform the design of the tutorials. First, to enhance education outcomes, the tutorials should be hands-on and interactive \cite{AI:education:2, AI:education:6, AI:education:4, AI:education:5}, hopefully with immediate feedback, i.e., learners can manipulate data and observe changes in model predictions \cite{AI:education:6}. For example, after one attempts to remove gender bias from the training data, the model should update accordingly and generate less biased predictions. 

Second, to simulate real-world AI development and debiasing practices, it is important to expose AI creators to technical knowledge of AI gender bias (e.g., sources of gender bias, debiasing techniques), AI concepts (e.g., feature, label, model) and AI development components (e.g., dataset, code). This equips developers to apply the knowledge learned to real-world development. With this in mind, we intentionally distinguished our tutorials from existing materials customized for AI novices (e.g., \cite{bias:education:5}).

Third, to engage learners, the education should be themed around real-life scenarios, inspired by situational learning \cite{shih2021learning}. We chose two scenarios, i.e., autocomplete and recruitment, which learners regularly use or come across in their daily life, and are commonly seen with gender bias \cite{pena2020bias, noble2018algorithms}. 

\subsection{What To Teach: Drawing from Literature}

We relied on existing research in understanding, identifying, and mitigating gender bias in AI to design the content of our tutorials. With our tutorials, we aim to equip AI creators with \textit{awareness} of gender bias, including understanding and identifying gender bias \cite{bias:education:1} and being willing to address this issue \cite{leavy2018gender}, since programmer bias was one of the key contributing factors to gender bias in AI systems \cite{gender:bias:2}.  

\textit{Knowledge}-wise, we aim to teach AI creators sources of gender bias and debiasing techniques. Friedman and Nissenbaum categorized bias into \textit{pre-existing social bias}, which has its roots in social institutions, practices, and attitudes, \textit{technical bias}, which arises from the resolution of technical issues in the design, and \textit{emergent social bias}, which emerges only in a context of use \cite{friedman1993discerning}. The source of AI gender bias taught in our tutorials is mainly pre-existing social bias \cite{friedman1993discerning}, which was introduced into AI through training data \cite{gender:bias:2}, but we also wanted to convey the idea that technical bias could arise if bias was not well solved by AI creators. Debiasing techniques, on the other hand, could help AI creators mitigate gender bias from AI. Numerous debiasing techniques have been proposed \cite{gender:bias:2, gender:bias:7}. Under the broad umbrella of data manipulation \cite{gender:bias:7}, one can, for example, debias the training corpora with data augmentation, e.g., to create an augmented dataset identical to the original dataset but biased towards the opposite gender (gender swapping) and to train on the union of the original and gender-swapped datasets \cite{gender:bias:8}. The two debiasing techniques (i.e., data cleaning and gender swapping) taught in our autocomplete tutorial and the first debiasing technique (i.e., data cleaning) taught in our recruitment tutorial could be categorized as data manipulation techniques. The second debiasing technique, unawareness, i.e., neglecting sensitive features such as gender, race, and age in the training process, taught in our recruitment tutorial, is a common practice in the machine learning domain \cite{gender:bias:11}.

\subsection{Two Tutorials: Recruitment and Autocomplete}
\begin{figure}[htbp]
    \centering
    \includegraphics[width=0.8\textwidth]{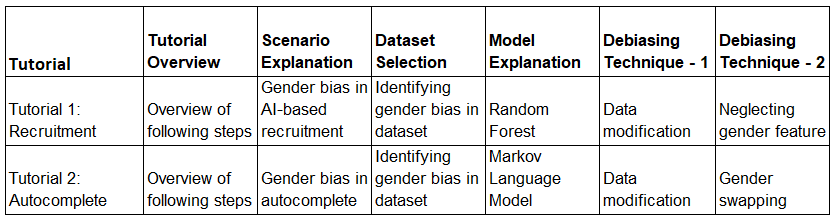}
    \caption{Tutorial user flow.}
    \label{tutorial:flow}
\end{figure}

The two tutorials both teach about sources of gender bias and debiasing techniques, based on two different scenarios. The user flow for the two tutorials is shown in Figure \ref{tutorial:flow}. Screenshots of the Recruitment Tutorial are shown in Figure \ref{tutorial:screenshot:1} and Figure \ref{tutorial:screenshot:2} in Appendix.

\subsubsection{The Recruitment Tutorial}
The recruitment tutorial aims to teach learners about gender bias in AI-based recruitment systems \cite{pena2020bias} -- when solely relying on AI for recruitment, fewer women may be hired, even if they have the same merits as men, due to the historical hiring prejudice toward women, especially in the STEM fields \cite{glass2010recruiting}. 

After an overview of the tutorial (Figure \ref{recruitment-overview}), the concept and harm of gender bias as well as gender bias in the hiring process are introduced (Figure \ref{recruitment-hiring}). Users are then prompted to choose between two datasets for training, one with gender bias, where well-performing women are not recruited, and one without (Figure \ref{recruitment-dataset}). After briefly explaining to users how Random Forest works, they are directed to Google Colaboratory\footnote{\url{https://colab.research.google.com/?utm_source=scs-index}. Google Colaboratory allows people to write and execute Python code through the browser.}, or Colab, which hosts the Python script for the models prepared by the research team  (Figure \ref{recruitment-random-forest}). They are instructed to run the models trained on the two different datasets and compare their hiring decisions on a new set of candidates. The model trained on the biased dataset will generate more biased predictions. The first debiasing technique, i.e., data cleaning, is then introduced (Figure \ref{recruitment-debiasing-1}). In this step, users are guided to clean the more biased dataset to make it less biased against women, use the debiased dataset to train the model, and check if there is bias in predictions. In the following step, users are presented with the second debiasing technique, i.e., neglecting gender feature in training, in a similar way (Figure \ref{recruitment-debiasing-2}). 

\subsubsection{The Autocomplete Tutorial}
The autocomplete tutorial aims to teach learners about gender bias in autocomplete systems which are commonly used in search engines. We hope to show users that stereotypical words, and specifically professional titles, are associated with different genders. For example, the word ``nurse'' has a significantly higher chance to be associated with women whilst ``doctor'' associated with men \cite{garrido2021survey}. Given such stereotypes, autocompleted search queries can lead to biased results for people.

The tutorial first introduces the definition and examples of autocomplete: ``If one types `Today,' AI might predict `is a good day,' since this is a frequent pattern in people's expressions.'' Then we present a biased, text-based dataset of gender-profession pairs (e.g., ``woman nurse'', ``man engineer'') where different genders have different distributions of profession. Then users are introduced to the Markov Language Model and prompted to observe predictions of the model trained on the biased dataset. Users are then asked to modify the dataset to make it less biased and observe the corresponding model predictions. The second debiasing technique, gender swapping, i.e., swapping the professions for men and women, is similarly taught. 

\subsection{Prototyping and Implementation}
We used Figma\footnote{\url{https://www.figma.com/}}, a UI prototyping tool, to design the Lo-Fi and Hi-Fi prototypes. In the Lo-Fi prototype, we designed the general layout and workflow of the websites. In the Hi-Fi prototype, we designed specific UI components, such as text, buttons, visualizations of models, and embedded links to datasets and Colab notebooks hosted on Google Drive. The websites were built with React\footnote{\url{https://react.dev/}} and Gatsby\footnote{\url{https://www.gatsbyjs.com/}} frameworks and deployed using Netlify\footnote{\url{https://www.netlify.com/}}. 
The recruitment tutorial\footnote{\url{https://recruitment-lab.netlify.app/}} and the autocomplete tutorial\footnote{\url{https://autocomplete-lab.netlify.app/}} are made publicly available.

\subsection{Pilot Study}
We conducted a pilot study with 5 participants, asking them to use the implemented tutorials and provide feedback on education content and design features. The pilot findings informed our design iteration and experiment process. First, a preview of the tutorial content and a progress bar were suggested by the participants to help them navigate the tutorials with more confidence, which we implemented in the final tutorials. Second, to improve the engagement and effectiveness of the tutorials, we increased the complexity of the datasets and provided more education about technical knowledge such as how the AI models work, as expected by the pilot participants who had relatively more tech background. We avoided jargon in the explanations to prevent people with less tech backgrounds from being intimidated. Lastly, since the participants consistently expressed that the autocomplete tutorial was harder to understand than the recruitment tutorial, in the later formal study, we presented the recruitment tutorial first to create a smoother learning curve for the participants.
\section{Method}
We conducted user experiments with 18 AI creators, either researchers, industrial practitioners, or students with an AI background, to evaluate the effectiveness and user experience of our AI gender bias tutorials. 

\subsection{Participants}

We recruited participants by posting messages and posters on Twitter and WeChat through a convenience sampling approach \cite{convenience}. We also asked our contacts who were CS students, CS professors, or AI developers, to spread the recruitment information. The screening survey (see more details in Section \ref{screening} in Appendix) asked about respondents' profession, major (for students), self-rated level of knowledge in AI, and 6 knowledge questions (KQs). Our recruitment started on June 21st and ended on July 4th, 2023, with a total of 1,382 responses collected. We screened for people who had prior experience with AI and who were over 18 years old. In the end, we selected 18 AI creators who had prior experiences with AI, including 8 AI researchers (4 studying AI and 4 studying the intersection between AI and HCI), 5 industrial practitioners (3 AI developers and 2 AI product managers), and 5 students who majored in computer science or had taken AI courses. The participants were equally distributed between female and male, with most of them from the US and China. Courses are one of the main sources of AI knowledge for our participants (16/18). Demographic information of the participants can be seen in Table \ref{demographic}. We compensated the participants with 100 RMB or 15 US Dollars for their participation in the user experiment which took place in Zoom for roughly 1 hour. The study is IRB-approved.

\begin{table}
\caption{Overview of participants.}
\label{demographic}
\centering
\begin{tabular}{lllll}
\textbf{Alias} & \textbf{Occupation} & \textbf{Gender} & \textbf{Country} & \textbf{Sources of AI Knowledge} \\
P1 & AI/HCI Researcher & F & China & Social media \\
P2 & AI/HCI Researcher & F & UK & Courses \\
P3 & AI Researcher & F & Germany & Courses, Social media \\
P4 & AI Developer & M & US & Courses, Social media, News \\
P5 & AI Product Manager & F & China & Courses \\
P6 & AI Researcher & M & US & Courses \\
P7 & AI Product Manager & F & Netherlands & Courses \\
P8 & AI Researcher & F & US & Courses, Social media, News, Research \\
P9 & AI Developer & M & US & Courses \\
P10 & AI Developer & M & US & Courses, Social media, News, Work \\
P11 & AI/HCI Researcher & F & China & Courses, Social media, News \\
P12 & AI/HCI Researcher & M & China & Courses, Social media, News, Books \\
P13 & Undergrad Student & M & China & Courses, Social media, News \\
P14 & Undergrad Student & M & China & Research \\
P15 & Undergrad Student & M & China & Courses, Social media, News \\
P16 & Undergrad Student & F & US & Courses, Social media \\
P17 & Undergrad Student & F & US & Courses \\
P18 & AI Researcher & M & China & Courses, Research
\end{tabular}
% \\hline
\end{table}

\subsection{Procedures}
We first introduced the purpose of the study and the study process. After acquiring consent and permission for recording from the participants, the study began. The experiment protocol is attached in Section \ref{protocol} in Appendix.

\subsubsection{Pre-Study Interview}
We asked participants a set of questions to probe their experience and perception of AI and gender bias, e.g., their experience of coming across gender bias in AI use and their perceived reasons for gender bias in AI. Given the educational nature of our study and our participants' CS background, we further asked if AI gender bias was taught in CS/AI courses and their debiasing practices when creating AI systems.

\subsubsection{Study Tasks: Completing Two Tutorials}
Participants were asked to complete two hands-on tutorials. During the learning process, we asked questions about their in-activity behaviors, e.g., ``Why did you choose this dataset?'' ``How do you think about this debiasing technique?'' 

After each tutorial, the participants were asked to fill in a survey, which asked the same six KQs as in the screening survey, e.g., choosing scenarios containing gender bias, choosing AI training phases where gender bias can be introduced, identifying biased training data, and choosing debiasing methods. We further asked Likert-scale usability questions, e.g., ``The tutorial was well organized and made good use of time,'' ``I learned about something technical,'' and ``I learned about something important for society,''

\subsubsection{Post-Study Interview}
After completing two hands-on tutorials, participants were asked about the experiment process and their understanding of AI gender bias. We particularly sought to see if their understanding changed after the education, e.g., ``How is gender bias introduced in the training and deployment process? Has your understanding changed after
our experiment?'' ``How can gender bias be mitigated from AI systems from a developer's perspective? Has your understanding
changed after our experiment?'' We ended by asking for suggestions to further improve the tutorials.

\subsection{Measurements}
We drew on several measurements to assess the educational effect and usability of the tutorials. Education effect was inferred through (1) participants' self-expressed awareness (ability to identify gender bias, willingness to mitigate gender bias) and knowledge (sources of gender bias, debiasing techniques) in the pre- and post-study interviews and (2) KQs in the surveys filled before and after the tutorials. The usability of the tutorials was inferred from (1) the post-study interview and (2) the usability questions in the surveys after the tutorials.

\subsection{Data Analysis}

\subsubsection{Qualitative Analysis}
We conducted a thematic analysis \cite{thematic} to understand the qualitative data, including transcripts of the pre-study and post-study interviews and observational notes taken during the experiments. Three authors independently and iteratively conducted open coding throughout the project. We regularly discussed, created/re-created codes, and identified emergent themes. We stopped coding when we observed
signs of saturation, i.e., no new themes emerged to answer our RQs. XMind\footnote{\url{https://xmind.app/}}, a mind-mapping tool, was used to organize themes, sub-themes, and quotes into a hierarchical structure. The major themes identified in the end included ``lived experience of prevalent AI gender bias,'' ``lack of AI gender bias education,'' ``awareness change,'' ``knowledge gain,'' ``intuitiveness,'' and so on. 

\subsubsection{Quantitative Analysis}
A quantitative analysis was applied to the KQ surveys answered before the study and after each tutorial to see if there was an improvement. We also analyzed the responses to the usability questions to understand the user experience of the tutorials. Descriptive statistics of KQ correctness and usability scores are calculated and reported in the next Section.
\section{Findings}
Our interviews revealed a lack of AI gender bias education in CS/AI courses the participants took, which led to a lack of awareness and knowledge of gender bias in them and their lack of confidence in debiasing AI. 
Our hands-on tutorials effectively improved the AI creators' awareness and knowledge of AI gender bias. After completing the tutorials, the participants had a more in-depth understanding of sources of gender bias and debiasing techniques. They were more inclined to pay attention to debiasing AI in their future development practices. Participants perceived our hands-on tutorials as intuitive and informative. 

\subsection{Insufficient Education Despite Prevalent AI Gender Bias (RQ1)}
We answered RQ1 based on findings collected from the pre-study interview to understand the participants' experience and perception of AI gender bias, and how this topic was covered in CS/AI courses. While they mentioned the wide existence of gender bias in AI they used or built, insufficient education regarding this issue was in place in the CS/AI courses they took.

The industrial practitioners provided inner observations of gender bias in the AI systems they built, especially regarding the difficulty of mitigating gender bias due to other corporate priorities and the male-dominated IT culture. P7 was the product manager of a Bert-based \cite{koroteev2021bert} educational chatbot for teaching English to children in China. She acknowledged that the design of the chatbot contained gender bias and stereotypes --- the chatbot was designed with a female figure and a feminine voice, just like many other voice assistants on the market \cite{feine2020gender}. As a feminist, she proposed to redesign the ``female'' chatbot with a more adventurous character to make the design less stereotypical, but was rejected by her manager given other corporate priorities and the manager's lack of gender awareness: 
\begin{quote}
    \textit{``Chatbots are always designed as caring and gentle female characters, which I suppose is a product of the male gaze and a way to attract male users. I wanted to make it less stereotypical, which I guess was not the priority of the team. The priority, instead, was to ship the product fast. It was a startup company and profit is all that is cared about. My manager doesn't think it is gender bias, and she doesn't care.''} 
\end{quote}
P10 was the recommendation algorithm manager for a well-known short video-sharing platform. He disclosed that their system took gender as an important feature when recommending content to users. Further, the cold start process, i.e., establishing user profiles and pushing recommendations for new users, assumed users as male for efficiency. He commented that this was a standardized way of building recommendation systems in the social media industry, possibly because ``most IT managers and developers were male and did not have adequate awareness of gender bias.''

Despite the perceived prevalence and severity of gender bias in AI systems, the participants regarded education on this topic as insufficient in CS/AI courses, where most of them were trained. Some noted that the discussion of gender bias as well as other ethical issues was completely absent from AI courses. Instead, only algorithms were taught. P2, who obtained her bachelor's degree in the UK, explained: 
\begin{quote}
    \textit{``AI and CS courses didn't teach bias or other ethical concerns at all but only taught principles and applications of AI algorithms. The assignments didn't concern gender bias as well.''}    
\end{quote}

Even if some reported receiving AI gender bias education in CS/AI courses, it was rather sporadic, appearing as a brief discussion in class. For example, P5, who received her postgraduate education in the UK, recalled that gender bias and other social and ethical issues were only occasionally taught in AI courses, appearing as quick discussions.
Similarly, only one slide was devoted to discussing AI gender bias throughout P9's CS/AI education in college.

Furthermore, AI gender bias was often discussed abstractly, e.g., under the umbrella of more generic bias and fairness principles. For instance, P8 recalled that in a foundational machine learning course, the instructor taught how data imbalance, e.g., more images of one demographic group than another in a dataset, led to fairness issues in AI models, and then briefly mentioned AI gender bias as one of the cases. P3, alike, noted that AI textbooks only discussed definitions and metrics of AI bias, but did not discuss specific examples concerning gender or race.

Yet another drawback of AI gender bias education in CS/AI courses was that it was often tech-oriented without engaging discussions about people's experiences and the relationship between people and technology, demonstrating a technological determinism perspective \cite{drew2016technological}. P9 gave an example,
\begin{quote}
    \textit{``The ethical discussions didn't touch on social aspects, but only focused on data processing, algorithms, and training infrastructures.''}
\end{quote} 
P7 mentioned an assignment about detecting prejudicial speech against women, which only asked for a higher accuracy of the model without facilitating a discussion around gender bias or data feminism \cite{data:feminism}. The only exception was an AI ethics course at Stanford P4 attended. He elaborated, 
\begin{quote}
    \textit{``The focus of the course was to remind you of the existence of AI bias, not the debiasing methods. Not many technical skills were taught in the course. The discussions were more about reflection of social issues.''}
\end{quote}

In short, education on AI gender bias was insufficient in the CS/AI courses the participants took. The discussion of AI gender bias in the classroom was absent, sporadic, too abstract, or tech-oriented. The lack of a comprehensive AI gender bias education combining technical knowledge and sociotechnical discussion was a missed opportunity to equip students with gender awareness and knowledge to combat gender bias in AI. The lack of education on AI gender bias may be attributed to insufficient gender awareness in CS professors who do not have a sociotechnical mindset, an emphasis on practical technical knowledge, fewer jobs related to AI ethics than software engineering, and so on. These are our speculations which can only be confirmed by interviewing course instructors. Nevertheless, by revealing the lack of education on AI gender bias, we demonstrate the necessity of delivering such education through educational tools and materials.

\subsection{Learning Outcomes (RQ2)}

Both awareness change and knowledge gain were observed through the participants' self-expressed improvement, evidence in KQ surveys, and their in-tutorial activities. Below, we unpack both aspects.

\subsubsection{Awareness Change: An Ability to Recognize Gender Bias and Willingness to Solve This Issue} 
Gender bias is often implicit and hard to recognize; making well-informed decisions about gender bias requires gender awareness and consciousness \cite{gender:bias:12}. Our tutorials helped participants gain awareness of AI gender bias in that they were more capable of understanding and identifying gender bias, more mindful of the harm gender bias could bring to women, and more willing to solve this issue.

\textbf{Recognizing Gender Bias.} In the pre-study interview, many participants were not aware of gender bias in AI. They realized it only after being nudged by the researchers (e.g., ``Are you recommended disproportionately more short videos about makeup than sports or tech? Do you think it's gender bias?''). Even P3, who self-identified as a feminist, indicated the difficulty of identifying gender bias in AI due to its implicit nature: 
\begin{quote}
    \textit{``Gender bias is often implicit. I read many books on feminism, so I can recognize gender bias, but not always. Most people would think it's just normal.''} 
\end{quote}
Similarly, P15 thought large language models (LLMs) output gender bias in a subtle, implicit way (e.g., assuming a nurse is female), and many people, including P15 himself, would not even realize that it was gender bias. Gradually, people would take gender bias for granted and lose the ability to identify gender bias, according to him. 

After completing the tutorials, P4 commented that he was able to recognize gender bias in different forms such as stereotyping (gender-profession association) and disparagement (recruitment decision based on gender). Participants commonly expressed such an awareness change. Participants' enhanced ability to recognize gender bias was observed in their in-tutorial activities. All participants but one were able to identify gender bias and select the less biased datasets while completing the tutorials. P1 provided her rationale for selecting the less biased dataset in the recruitment tutorial, which was common among other participants: ``In the other dataset, women are not recruited even if they perform well in interviews.'' P5 made the wrong selection of dataset in the recruitment tutorial since she was not familiar with data sheets and did not spot any differences between the two datasets. The participants' ability to recognize gender bias was further evidenced by an increased accuracy of the KQ which asked them to choose the scenarios that contained gender bias. Five participants out of 18 made the correct choices before the study. After learning the recruitment tutorial, 8 of them answered the question correctly. The headcount rose to 9 after the autocomplete tutorial.

\textbf{Willingness to Solve AI Gender Bias.} None of our participants had professional experience debiasing AI systems, due to a lack of consideration of the harm gender bias could bring, or a focus on non-human systems such as robots. P17 thought our tutorials prompted her to think about how gender bias affected the career and browsing experience of women, which she tended to overlook in her life. Such ignorance of the existence and harm of AI gender bias was more common and severe in male AI creators. For example, P12 deemed gender bias as less harmful and offensive. P10 told the concerning fact that many male engineers in his development team were not equipped with adequate gender awareness and would ``laugh about feminism when talking about the \#MeToo Movements.'' P6 frankly acknowledged that he did not care much about AI gender bias since it did not impact him negatively. Since gender bias often targeted women and gender minorities, he suggested that he could even ``benefit from gender bias as a man such as enjoying better work opportunities in the IT field.'' As a result, he did not consider debiasing AI when he built AI systems.

Our participants expressed a stronger intention to pay attention to debiasing AI systems after completing the tutorials. The mindset of P6, who indicated little interest in debiasing AI, changed -- after completing the tutorials, he realized the harm AI gender bias could incur to women, which he had not been exposed to as a man: 
\begin{quote}
    \textit{``Previously, I didn't think keeping recommending girls dolls was biased. Now, I realize that the current society has more flexible gender roles assigned to people, and we should do something to combat stereotypes in AI. They caused real harm to women.''} 
\end{quote}
P8, an AI researcher who was shifting her research focus from non-human systems to sociotechnical systems such as LLMs, indicated that she would consider gender bias: 
\begin{quote}
    \textit{``I didn't think much about gender bias before, since I've been working on robots. Now I find it important to use subjective, manual interference to solve gender issues in social contexts after seeing how severe gender bias can be in sociotechnical systems such as search engines which people use daily. I'll consider gender bias as I dive into the research of LLMs.''}
\end{quote} 
She suggested that the real-life scenarios enabled her to think about how gender bias impacted people.

Overall, through completing our tutorials, the participants showed increased awareness of gender bias in AI. They were more able to identify gender bias and more willing to mitigate gender bias from AI systems.

\subsubsection{Knowledge Gain: Sources of Gender Bias and Debiasing Techniques}

In the pre-study interview, most participants acknowledged a lack of knowledge in terms of sources of gender bias and debiasing techniques and were not confident about building bias-free AI systems, which may be attributed to insufficient school education on AI gender bias and education tools/materials. P2 said: 
\begin{quote}
    \textit{``I'm not confident in debiasing AI systems since I know too little about gender bias. Few people are equipped with this knowledge.''}
\end{quote}
P6 speculated that ``gender bias would probably remain'' if he were to build an AI system. Even if people had relevant knowledge, ``it was not easy to apply the theoretical knowledge into practice,'' as P3 commented. Below, we report their knowledge gain in terms of sources of gender bias and debiasing techniques.

\textbf{Sources of Gender Bias.} Before the study, some participants only had a simple understanding of the sources of gender bias. Pre-existing social bias was seen as a main source of gender bias in AI systems. For example, when we asked P5 to reflect on the Google autocomplete scenario in the screening survey, she thought gender bias rooted in society and human language was the reason for AI's misbehavior: 
\begin{quote}
    \textit{``Gender bias has existed in society for a long time. People's search behavior and language patterns are learned by AI. People pay more attention to women's body shape and personal lives and pay more attention to men's careers. AI simply reinforces such social biases.''} 
\end{quote}
P9, who had a more technical background, cited the classic ``garbage in, garbage out'' concept, meaning the quality of model output is determined by the quality of the input.

Our tutorials helped improve the participants' knowledge of sources of AI gender bias.
The tutorials corrected the false understanding of some participants, as in P3's case:
\begin{quote}
    \textit{``Before the study, I thought AI was a black box and maybe it set some rules to contain bias in itself. Now I know gender bias is introduced by the datasets which carry social biases.''}
\end{quote}
Some gained a more comprehensive understanding of sources of gender bias through the tutorials. For example, P2 originally thought the only reason for biased AI was ``social contexts'' (pre-existing social bias). After completing our tutorials, she realized that gender bias also came from the training process when developers made wrong decisions to handle gender bias (technical bias) and that ``one could mitigate gender bias to some extent if they made some efforts.'' 

\textbf{Debiasing Techniques.} The participants with less technical knowledge of AI, like AI/HCI researchers and AI product managers, had never heard of debiasing techniques before the study. The AI researchers and AI developers who had heard of debiasing techniques did not know how they worked or how to apply them in practice. 

After the study, our participants reported a deeper understanding of debiasing techniques in AI systems and felt more natural and confident doing so. For example, P2 did not know any debiasing techniques before the study and thought debiasing was a hopeless effort. She was surprised by the power of debiasing techniques after mitigating gender bias from AI models by herself in the tutorials. Similarly, P11 originally thought gender bias in datasets was hard to remove, but gained more experience and confidence after practicing the debiasing techniques.
P1's mindset switched from ``human biases are hard to solve'' to ``I can solve the problem after learning and practicing the debiasing techniques.'' In the KQ survey, more participants chose the debiasing techniques correctly (pre-study: 4/18, after recruitment tutorial: 9/18, after autocomplete tutorial: 10/18).

Some participants expressed a persistent challenge in the development of bias-free AI -- debiasing was often not a priority for AI companies, so they may not be able to apply the debiasing techniques they learned. For example, P9 argued that unawareness, i.e., neglecting the gender feature in training, might not be feasible because ``asking companies to give up the collected gender feature is not very likely unless it is enforced by policy or law.'' P10 thought there were no direct financial benefits to incentivize companies to mitigate gender bias. Instead, recommending biased content can even maximize profits in some cases. He explained, 
\begin{quote}
    \textit{``When most users are male, debiasing would lead to poorer performance for users, with core metrics degraded. Recommending content with beautiful women to male users maximizes financial returns for social media platforms... Since debiasing was not a corporate priority, no fairness-related values are coded into the mindset of developers.''}
\end{quote}
P4 pointed out the massive cost of debiasing AI since ``companies need to hire people who have this knowledge.'' 
P13 and P18 noted that developers were not often the decision-makers. Even if they were willing and able to debias AI models, ``they can only do what they are told. (P13)''

Overall, after our tutorials, the participants showed increased knowledge of sources of gender bias and debiasing techniques. With this knowledge, they exhibited more confidence in debiasing AI systems, though other corporate priorities may hurdle their debiasing practices.

\subsection{User Experience of Our Hands-on Tutorials (RQ3)}

\begin{figure}[htbp]
    \centering
    \includegraphics[width=0.8\linewidth]{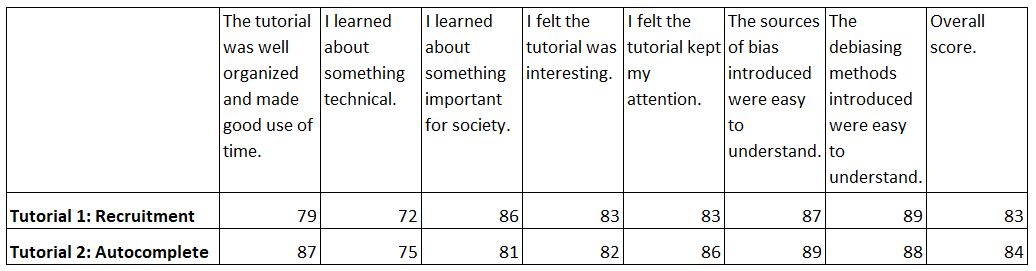}
    \caption{Usability scores summary - average evaluations. Converted from 7-point Likert Scale.}
    \label{usability:score}
\end{figure}

The participants rated our tutorials highly in terms of: usability; being interesting, engaging, and easy to understand; and teaching topics that are important for society. Based on the usability questions (see Section \ref{protocol} in Appendix), the participants rated the recruitment tutorial 83 out of 100, and the autocomplete tutorial 84 out of 100, indicating approval. Both tutorials did relatively poorly in one area -- delivering technical knowledge (72 for the recruitment tutorial, 75 for the autocomplete tutorial), which could be attributed to our design considerations. First, we wanted to strike a balance between technical knowledge and sociotechnical discussion. Second, we wanted to accommodate different levels of technical background of AI creators. The overall usability score and those for different aspects can be seen in Figure \ref{usability:score}.

Participants particularly valued the hands-on nature of the tutorials, which made the education more intuitive and engaging. P5 found learning AI gender bias through manipulating data and observing the impact on model predictions to be intuitive. P3 elaborated: 
\begin{quote}
    \textit{``After the study, I have a more concrete understanding of gender bias and debiasing methods. I'm curious to see the new results after modifying the datasets. The hands-on nature keeps researchers and participants on the same page and makes the discussion more concrete.''}
\end{quote} 
During the learning process, all participants were able to operate and learn from the key tasks such as selecting the less biased dataset, modifying the more biased dataset, and applying different debiasing techniques. 

The tutorials equipped tech people like AI developers and AI researchers with social awareness, which was often lacking in them. P6 noted:  
\begin{quote}
    \textit{``The tutorials enable me to realize hidden bias through simple demos. This is especially important for technical people like me.''} 
\end{quote}

The participants also expressed several issues and challenges they encountered when taking our tutorials. First, identifying differences between datasets was a time-consuming task, especially for people without a tech background. P8 suggested a visualized comparison of the datasets to show their discrepancies. Second, the autocomplete tutorial was perceived harder to grasp for many participants, since more domain knowledge was required to understand NLP models. Third, some participants suggested bringing racial bias into the discussion since it was closely related to gender bias. We elaborate more on these drawbacks and potential improvements in the Discussion section.
\section{Discussion}
Designing and evaluating the tutorials led us to three main findings.

\textbf{Finding 1: AI gender bias education is urgently needed yet insufficient.} There has long been feminist critique of AI systems \cite{adam1995feminist}, which are full of gender bias \cite{gender:bias:3, gender:bias:4, gender:bias:5, gender:bias:6}. Our participants spoke about biased recommendations by social media and shopping websites, prejudiced decisions made by AI-based recruitment systems, and so on. 

AI creators, who are responsible for debiasing AI, have been found to lack knowledge of preventive measures against AI bias, demonstrating a need to strengthen their education \cite{gender:bias:14}. Male AI creators have less awareness of gender bias than female AI creators \cite{leavy2018gender, medel2017eliminating}. Through the interviews, we similarly revealed a lack of awareness and knowledge of AI gender bias in a wide range of AI creators such as AI researchers, developers, and product managers. None of the participants were confident in building bias-free AI systems due to their lack of knowledge and corporate priorities such as expediency. 

Despite the urgent need for AI gender bias education, especially for AI creators who are on the front line of building bias-free AI, limited education efforts were seen. Contrary to the findings in \cite{gender:bias:8}, i.e., bias was a frequent topic in the 51 CS/AI courses they examined, our participants revealed that most CS/AI courses only taught AI models and their applications without touching on gender bias and other ethical concerns. Even if AI gender bias was mentioned, it was brief, abstract, and tech-oriented. Outside of the classroom, there is only one education tool for teaching AI gender bias to our knowledge \cite{bias:education:1}. This tool is designed for children, aiming to raise their awareness of gender bias through the explanation of stereotypical predictive features (e.g., predicting a person as female based on kitchen elements). It is not suitable for AI creators who need more technical knowledge such as debiasing techniques \cite{gender:bias:14}. A comprehensive gender bias education combining technical knowledge and sociotechnical discussions is needed.

\textbf{Finding 2: Our tutorials improved AI creators' awareness and knowledge of AI gender bias.}
Toward filling the vacuum of AI gender bias education for AI creators, we developed hands-on tutorials focused exclusively on this topic, complementing limited existing efforts \cite{bias:education:1}.

Our tutorials improved the awareness and knowledge of AI gender bias for AI creators. Awareness-wise, our participants felt more confident in identifying gender bias and more willing to solve the bias issue after completing the tutorials. Knowledge-wise, they were equipped with knowledge of sources of gender bias and debiasing techniques, which was key to debiasing AI \cite{gender:bias:14}.

\textbf{Finding 3: Hands-on activities are highly effective in engaging learners.}
Inspired by AI ethics education tools designed for AI novices \cite{bias:education:5, bias:education:3, bias:education:2}, we designed the tutorials to be hands-on and scenario-based and provide immediate feedback, making them intuitive and engaging for learners. Even those with limited AI knowledge could benefit from the learning process by manipulating the dataset, running the model, and observing the predictions. The real-life scenarios, recruitment and autocomplete, echoed the participants' life experiences and enabled them to reflect on the harm gender bias could incur. 
                                 
\subsection{Education Implications: A Rubric}

Our AI gender bias education is just a start; by accommodating intersectional experiences, education on AI biases in general can be similarly approached. To ease future efforts in delivering AI bias education, we provide a rubric (see Table \ref{rubrics}), outlining what to teach and how to evaluate the education. 

People's assumptions about gender are often skewed -- an online study showed that college students on average preferred the original biased recommendation of college majors over the debiased recommendation \cite{gender:bias:12}. The same can be said about people's racial assumptions -- associations between leaders and Whiteness held up to scrutiny \cite{petsko2023leaders}. To equip people with more adequate awareness and assumptions of gender, race, and other identities, and enhanced ability to identify bias, in the context of AI in particular, an ideal education should teach them to understand and identify bias in AI, as well as motivate them to do so. At the knowledge level, one can be taught sources of bias (social bias and technical bias) and debiasing techniques (dataset modification, model modification). Even AI creators are found to lack such knowledge \cite{gender:bias:14}, let alone users who have less technical knowledge of AI.

In Table \ref{rubrics}, we list learning objectives for education in terms of awareness and knowledge, respectively, as well as tasks, evaluation methods, and rating criteria, which are subject to change in different contexts. For example, our tutorials can be adapted into hands-on group activities accompanied by in-class discussions in a high school classroom \cite{bias:education:7}. Researchers and teachers can use this rubric as a guide when developing educational materials. 

\begin{table}
\caption{A rubric for teaching AI bias.}
\label{rubrics}
\begin{center}
\begin{tabular}{ c c c c c }
 \textbf{Category} & \textbf{Learning Objective} & \textbf{Tasks} & \textbf{Evaluation} & \textbf{Rating} \\ 
Awareness & Understanding & Defining Bias & Tutorial & Yes/No \\  
  & Identifying & Dataset Selection & Tutorial & Yes/No \\  
  & Willingness to Address & Self-Reporting & Exit Interview & Yes/No \\  
 Knowledge & Sources of Bias & Social Bias & Exit Interview & (0-10) \\
  &  & Technical Bias & Exit Interview & (0-10) \\
 & Debiasing Techniques & Dataset Modification & Tutorial, Exit Interview & (0-10) \\
  & & Model Modification & Tutorial, Exit Interview & (0-10) \\
\end{tabular}
\end{center}
\end{table}

\subsection{Design Implications}
\label{lessons}

Here we share implications for the design and evaluation of future AI gender bias education tutorials.

\subsubsection{Equipping Tech People with Gender Awareness}
Tech people, especially AI researchers studying non-human systems and AI developers, often lack awareness of AI gender bias, as seen in our participants and results in previous research \cite{gender:bias:14}. Further, our female participants expressed more concern about AI gender bias given their negative experience with AI, while some male participants even claimed to have ``benefited from gender bias.'' This echoed with prior research on the gendered perception of AI biases 
\cite{gender:bias:14}. Thus, it is important to equip tech people with gender awareness, so that they can mindfully mitigate gender bias in AI. Our tutorials successfully educated AI creators, who previously had limited or skewed awareness of AI gender bias, on gender bias in AI and motivated them to mitigate it.

\subsubsection{Accommodating Different Levels of Tech Literacy}
Considering how developmental milestones and prior experience with technology affect perceptions of AI is important when developing educational tools. Much effort has been devoted to communicating AI concepts to learners without extensive backgrounds in math or CS, including children \cite{AI:education:4, AI:education:5}. However, the simplified communication of AI concepts \cite{bias:education:1} may not satisfy the learning needs of AI creators, who need to deal with more complicated scenarios when developing AI systems. By exposing learners to data sheets, code, and the training process with sufficient explanation, while at the same time avoiding the need to write code and interpret tech jargon, we accommodate the learning needs of both AI novices and more tech-savvy learners. 

\subsubsection{Promoting Workforce Diversity}
Gendered attitudes toward AI gender bias were found in AI developers \cite{gender:bias:14}, i.e., female AI developers pay more attention to pursuing gender neutrality and reducing gender stereotypes when designing AI. However, they were often not the decision-makers in the male-dominant industry \cite{gender:bias:15, leavy2018gender}. Mitigating the gender gap in the computing workforce has the potential to reduce gender bias in the development process \cite{herring2006gender}, though challenging the inequalities of the capitalist political-economic system and the patriarchal society we live in is extremely difficult and complicated  \cite{toupin2023shaping}. Democratizing AI education through readily available tutorials like ours seems a viable way to engage more women in the CS/AI field.

\subsection{Limitations and Future Work}

There are several limitations of our work. First, we looked at gender identity in isolation from other factors such as race \cite{gender:bias:10}. As a result, the tutorials did not prepare participants to identify the complex effects of membership in multiple marginalized communities. These intersectional effects are widely overlooked in the algorithmic fairness literature, but some work has shown that persons at the intersections of multiple marginalized identities may be distinctly at risk of algorithmic marginalization \cite{van2022intersectional}. Future work will need to develop education to prepare AI creators to assess intersectional effects, which may require special consideration given the sometimes complex and surprising ways in which those effects manifest.

Second, profit-oriented corporate priorities such as shipping products fast and ``profitable bias'' are persisting hurdles to debiasing AI. Future research could combine education with policy to push companies into the collective mission of debiasing AI \cite{hine2023blueprint}.

Third, our tutorials only covered the parts of the AI training lifecycle before deployment. In reality, good performance in training may not always lead to satisfactory performance after deployment. Teaching how gender bias can be introduced into AI systems after deployment and how continuous monitoring is necessary is important for AI creators to develop more bias-free AI. 

Last, we did not address the design suggestions expressed by the participants since our main goals were to validate the design concepts and answer our research questions. More research and design efforts are needed to make education materials and tools more effective and usable.
\section{Conclusion}
AI creators often lack the awareness and knowledge to mitigate gender bias from AI. We design and evaluate hands-on tutorials to complement the insufficient school education in AI gender bias and education tools. Evaluations with AI creators demonstrate the effectiveness of our tools -- the participants' awareness and knowledge increased after completing our tutorials. We reflect on the lessons learned during our design and evaluation process in terms of how our tutorials succeed and how they can be designed better. We call for more efforts in designing AI gender bias education tools and materials to prepare AI creators to build bias-free AI.

\section*{Acknowledgments}
We want to thank Bang An and Ted Underwood for their valuable feedback at the early stage of this paper.

\bibliographystyle{plain}
\bibliography{chi}

\begin{thebibliography}{10}

\bibitem{abrams2015gender}
Jessica~R Abrams, Amy~M Bippus, and Karen~J McGaughey.
\newblock Gender disparaging jokes: An investigation of sexist-nonstereotypical jokes on funniness, typicality, and the moderating role of ingroup identification.
\newblock {\em Humor}, 28(2):311--326, 2015.

\bibitem{adam1995feminist}
Alison Adam.
\newblock A feminist critique of artificial intelligence.
\newblock {\em European Journal of Women's Studies}, 2(3):355--377, 1995.

\bibitem{bias:education:2}
S.~Akgun and C.~Greenhow.
\newblock Artificial intelligence in education: Addressing ethical challenges in k-12 settings.
\newblock {\em AI and Ethics}, pages 1--10, 2021.

\bibitem{ali2019discrimination}
Muhammad Ali, Piotr Sapiezynski, Miranda Bogen, Aleksandra Korolova, Alan Mislove, and Aaron Rieke.
\newblock Discrimination through optimization: How facebook's ad delivery can lead to biased outcomes.
\newblock {\em Proceedings of the ACM on human-computer interaction}, 3(CSCW):1--30, 2019.

\bibitem{bias:education:3}
S.~Ali, B.~H. Payne, R.~Williams, H.~W. Park, and C.~Breazeal.
\newblock Constructionism, ethics, and creativity: Developing primary and middle school artificial intelligence education.
\newblock In {\em International workshop on education in artificial intelligence k-12}, pages 1--4, 2019.

\bibitem{AI:education:2}
K.~E.~K. Bilstrup, M.~H. Kaspersen, I.~Assent, S.~Enni, and M.~G. Petersen.
\newblock From demo to design in teaching machine learning.
\newblock In {\em ACM Conference on Fairness, Accountability, and Transparency}, pages 2168--2178, 2022.

\bibitem{borenstein2021emerging}
Jason Borenstein and Ayanna Howard.
\newblock Emerging challenges in ai and the need for ai ethics education.
\newblock {\em AI and Ethics}, 1:61--65, 2021.

\bibitem{thematic}
V.~Braun and V.~Clarke.
\newblock Using thematic analysis in psychology.
\newblock {\em Qualitative research in psychology}, 3(2):77--101, 2006.

\bibitem{broussard2023more}
Meredith Broussard.
\newblock {\em More Than a Glitch: Confronting Race, Gender, and Ability Bias in Tech}.
\newblock MIT Press, 2023.

\bibitem{AI:education:6}
M.~Carney, B.~Webster, I.~Alvarado, K.~Phillips, N.~Howell, J.~Griffith, and A.~Chen.
\newblock Teachable machine: Approachable web-based tool for exploring machine learning classification.
\newblock In {\em Extended abstracts of the 2020 CHI conference on human factors in computing systems}, pages 1--8, 2020.

\bibitem{cirillo2020sex}
Davide Cirillo, Silvina Catuara-Solarz, Czuee Morey, Emre Guney, Laia Subirats, Simona Mellino, Annalisa Gigante, Alfonso Valencia, Mar{\'\i}a~Jos{\'e} Rementeria, Antonella~Santuccione Chadha, et~al.
\newblock Sex and gender differences and biases in artificial intelligence for biomedicine and healthcare.
\newblock {\em NPJ digital medicine}, 3(1):81, 2020.

\bibitem{dastin2022amazon}
Jeffrey Dastin.
\newblock Amazon scraps secret ai recruiting tool that showed bias against women.
\newblock In {\em Ethics of data and analytics}, pages 296--299. Auerbach Publications, 2022.

\bibitem{de2023educating}
David De~Cremer and Devesh Narayanan.
\newblock On educating ethics in the ai era: why business schools need to move beyond digital upskilling, towards ethical upskilling.
\newblock {\em AI and Ethics}, pages 1--5, 2023.

\bibitem{dev2021measures}
Sunipa Dev, Emily Sheng, Jieyu Zhao, Aubrie Amstutz, Jiao Sun, Yu~Hou, Mattie Sanseverino, Jiin Kim, Akihiro Nishi, Nanyun Peng, et~al.
\newblock On measures of biases and harms in nlp.
\newblock {\em arXiv preprint arXiv:2108.03362}, 2021.

\bibitem{data:feminism}
C.~D'ignazio and L.~F. Klein.
\newblock {\em Data feminism}.
\newblock MIT press, 2020.

\bibitem{gender:bias:3}
A.~Domnich and G.~Anbarjafari.
\newblock Responsible ai: Gender bias assessment in emotion recognition.
\newblock {\em arXiv preprint arXiv:2103.11436}, 2021.

\bibitem{drew2016technological}
Rob Drew.
\newblock Technological determinism.
\newblock {\em A companion to popular culture}, pages 165--183, 2016.

\bibitem{eguchi2021contextualizing}
Amy Eguchi, Hiroyuki Okada, and Yumiko Muto.
\newblock Contextualizing ai education for k-12 students to enhance their learning of ai literacy through culturally responsive approaches.
\newblock {\em KI-K{\"u}nstliche Intelligenz}, 35(2):153--161, 2021.

\bibitem{ellemers2018gender}
Naomi Ellemers.
\newblock Gender stereotypes.
\newblock {\em Annual review of psychology}, 69:275--298, 2018.

\bibitem{convenience}
R.~W. Emerson.
\newblock Convenience sampling revisited: Embracing its limitations through thoughtful study design.
\newblock {\em Journal of Visual Impairment \& Blindness}, 115(1):76--77, 2021.

\bibitem{endo2018technological}
Seth~Katsuya Endo.
\newblock Technological opacity \& procedural injustice.
\newblock {\em BCL Rev.}, 59:821, 2018.

\bibitem{feine2020gender}
Jasper Feine, Ulrich Gnewuch, Stefan Morana, and Alexander Maedche.
\newblock Gender bias in chatbot design.
\newblock In {\em Chatbot Research and Design: Third International Workshop, CONVERSATIONS 2019, Amsterdam, The Netherlands, November 19--20, 2019, Revised Selected Papers 3}, pages 79--93. Springer, 2020.

\bibitem{friedman1993discerning}
Batya Friedman and Helen Nissenbaum.
\newblock Discerning bias in computer systems.
\newblock In {\em INTERACT'93 and CHI'93 Conference Companion on Human Factors in Computing Systems}, pages 141--142, 1993.

\bibitem{bias:education:8}
N.~Garrett, N.~Beard, and C.~Fiesler.
\newblock More than ``if time allows'': the role of ethics in ai education.
\newblock In {\em Proceedings of the AAAI/ACM Conference on AI, Ethics, and Society}, pages 272--278, 2020.

\bibitem{garrido2021survey}
Ismael Garrido-Mu{\~n}oz, Arturo Montejo-R{\'a}ez, Fernando Mart{\'\i}nez-Santiago, and L~Alfonso Ure{\~n}a-L{\'o}pez.
\newblock A survey on bias in deep nlp.
\newblock {\em Applied Sciences}, 11(7):3184, 2021.

\bibitem{glass2010recruiting}
Christy Glass and Krista~Lynn Minnotte.
\newblock Recruiting and hiring women in stem fields.
\newblock {\em Journal of diversity in Higher Education}, 3(4):218, 2010.

\bibitem{bias:education:6}
N.~Green.
\newblock An ai ethics course highlighting explicit ethical agents.
\newblock In {\em Proceedings of the 2021 AAAI/ACM Conference on AI, Ethics, and Society}, pages 519--524, 2021.

\bibitem{gender:bias:5}
M.~Gupta, C.~M. Parra, and D.~Dennehy.
\newblock Questioning racial and gender bias in ai-based recommendations: Do espoused national cultural values matter?
\newblock {\em Information Systems Frontiers}, pages 1--17, 2021.

\bibitem{herring2006gender}
Susan~C Herring, Christine Ogan, Manju Ahuja, and Jean~C Robinson.
\newblock Gender and the culture of computing in applied it education.
\newblock In {\em Encyclopedia of gender and information technology}, pages 474--481. IGI Global, 2006.

\bibitem{hine2023blueprint}
Emmie Hine and Luciano Floridi.
\newblock The blueprint for an ai bill of rights: in search of enaction, at risk of inaction.
\newblock {\em Minds and Machines}, pages 1--8, 2023.

\bibitem{gender:bias:6}
T.~Hitron, B.~Megidish, E.~Todress, N.~Morag, and H.~Erel.
\newblock Ai bias in human-robot interaction: An evaluation of the risk in gender biased robots.
\newblock In {\em 31st IEEE International Conference on Robot and Human Interactive Communication}, pages 1598--1605, 2022.

\bibitem{jacobson1992gender}
Jodi~L Jacobson et~al.
\newblock {\em Gender bias: roadblock to sustainable development.}
\newblock 1992.

\bibitem{karapapa2015search}
Stavroula Karapapa and Maurizio Borghi.
\newblock Search engine liability for autocomplete suggestions: personality, privacy and the power of the algorithm.
\newblock {\em International Journal of Law and Information Technology}, 23(3):261--289, 2015.

\bibitem{katznelson2021need}
Gali Katznelson and Sara Gerke.
\newblock The need for health ai ethics in medical school education.
\newblock {\em Advances in Health Sciences Education}, 26:1447--1458, 2021.

\bibitem{kazim2021high}
Emre Kazim and Adriano~Soares Koshiyama.
\newblock A high-level overview of ai ethics.
\newblock {\em Patterns}, 2(9), 2021.

\bibitem{bias:education:7}
Z.~Kilhoffer, Z.~Zhou, F.~Wang, F.~Tamton, Y.~Huang, P.~Kim, T.~Yeh, and Y.~Wang.
\newblock ``how technical do you get? i'm an english teacher''': Teaching and learning cybersecurity and ai ethics in high school.
\newblock In {\em 44th IEEE Symposium on Security and Privacy}, 2023.

\bibitem{koroteev2021bert}
MV~Koroteev.
\newblock Bert: a review of applications in natural language processing and understanding.
\newblock {\em arXiv preprint arXiv:2103.11943}, 2021.

\bibitem{lambrecht2019algorithmic}
Anja Lambrecht and Catherine Tucker.
\newblock Algorithmic bias? an empirical study of apparent gender-based discrimination in the display of stem career ads.
\newblock {\em Management science}, 65(7):2966--2981, 2019.

\bibitem{leavy2018gender}
Susan Leavy.
\newblock Gender bias in artificial intelligence: The need for diversity and gender theory in machine learning.
\newblock In {\em Proceedings of the 1st international workshop on gender equality in software engineering}, pages 14--16, 2018.

\bibitem{gender:bias:4}
M.~Makhortykh, A.~Urman, and R.~Ulloa.
\newblock Detecting race and gender bias in visual representation of ai on web search engines.
\newblock In {\em International Workshop on Algorithmic Bias in Search and Recommendation}, pages 36--50, 2021.

\bibitem{manyika2017future}
James Manyika, Michael Chui, Mehdi Miremadi, Jacques Bughin, Katy George, Paul Willmott, and Martin Dewhurst.
\newblock A future that works: Ai, automation, employment, and productivity.
\newblock {\em McKinsey Global Institute Research, Tech. Rep}, 60:1--135, 2017.

\bibitem{gender:bias:10}
T.~Manzini, Y.~C. Lim, Y.~Tsvetkov, and A.~W. Black.
\newblock Black is to criminal as caucasian is to police: Detecting and removing multiclass bias in word embeddings.
\newblock {\em arXiv preprint arXiv:1904.04047}, 2019.

\bibitem{medel2017eliminating}
Paola Medel and Vahab Pournaghshband.
\newblock Eliminating gender bias in computer science education materials.
\newblock In {\em Proceedings of the 2017 ACM SIGCSE technical symposium on computer science education}, pages 411--416, 2017.

\bibitem{bias:education:1}
G.~I. Melsión, I.~Torre, E.~Vidal, and I.~Leite.
\newblock Using explainability to help children understandgender bias in ai.
\newblock In {\em Interaction Design and Children}, pages 87--99, 2021.

\bibitem{mibenge2013sex}
Chiseche~Salome Mibenge.
\newblock {\em Sex and international tribunals: The erasure of gender from the war narrative}.
\newblock University of Pennsylvania Press, 2013.

\bibitem{gender:bias:1}
A.~Nadeem, O.~Marjanovic, and B.~Abedin.
\newblock Gender bias in ai-based decision-making systems: a systematic literature review.
\newblock {\em Australasian Journal of Information Systems}, 26, 2022.

\bibitem{gender:bias:2}
Ayesha Nadeem, Babak Abedin, and Olivera Marjanovic.
\newblock Gender bias in ai: A review of contributing factors and mitigating strategies.
\newblock In {\em ACIS 2020 Proceedings}, 2020.

\bibitem{noble2018algorithms}
Safiya~Umoja Noble.
\newblock Algorithms of oppression.
\newblock In {\em Algorithms of oppression}. New York university press, 2018.

\bibitem{pena2020bias}
Alejandro Pena, Ignacio Serna, Aythami Morales, and Julian Fierrez.
\newblock Bias in multimodal ai: Testbed for fair automatic recruitment.
\newblock In {\em Proceedings of the IEEE/CVF Conference on Computer Vision and Pattern Recognition Workshops}, pages 28--29, 2020.

\bibitem{petsko2023leaders}
Christopher~D Petsko and Ashleigh~Shelby Rosette.
\newblock Are leaders still presumed white by default? racial bias in leader categorization revisited.
\newblock {\em Journal of Applied Psychology}, 108(2):330, 2023.

\bibitem{gender:bias:11}
E.~Pitoura, K.~Stefanidis, and G.~Koutrika.
\newblock Fairness in rankings and recommendations: an overview.
\newblock {\em The VLDB Journal}, pages 1--28, 2022.

\bibitem{prates2020assessing}
Marcelo~OR Prates, Pedro~H Avelar, and Lu{\'\i}s~C Lamb.
\newblock Assessing gender bias in machine translation: a case study with google translate.
\newblock {\em Neural Computing and Applications}, 32:6363--6381, 2020.

\bibitem{quinn2021readying}
Thomas~P Quinn and Simon Coghlan.
\newblock Readying medical students for medical ai: The need to embed ai ethics education.
\newblock {\em arXiv preprint arXiv:2109.02866}, 2021.

\bibitem{gender:bias:15}
K.~Schulenberg, H.~Watkins, A.~I. Hauptman, E.~A. Schlesener, and G.~Freeman.
\newblock ``i felt like i wasn’t really meant to be there'': Understanding women's perceptions of gender in approaching ai design \& development.
\newblock In {\em Proceedings of the 56th Hawaii International Conference on System Sciences}, 2023.

\bibitem{bias:education:5}
B.~R. Shapiro, A.~Meng, C.~O'Donnell, C.~Lou, E.~Zhao, B.~Dankwa, and A.~Hostetler.
\newblock Re-shape: A method to teach data ethics for data science education.
\newblock In {\em Proceedings of the 2020 CHI conference on human factors in computing systems}, pages 1--13, 2020.

\bibitem{shih2021learning}
Po-Kang Shih, Chun-Hung Lin, Leon~Yufeng Wu, and Chih-Chang Yu.
\newblock Learning ethics in ai—teaching non-engineering undergraduates through situated learning.
\newblock {\em Sustainability}, 13(7):3718, 2021.

\bibitem{gender:bias:7}
T.~Sun, A.~Gaut, S.~Tang, Y.~Huang, M.~ElSherief, J.~Zhao, and W.~Y. Wang.
\newblock Mitigating gender bias in natural language processing: Literature review.
\newblock {\em arXiv preprint arXiv:1906.08976}, 2019.

\bibitem{tipler2019dehumanizing}
Caroline~N Tipler and Janet~B Ruscher.
\newblock Dehumanizing representations of women: the shaping of hostile sexist attitudes through animalistic metaphors.
\newblock {\em Journal of Gender Studies}, 28(1):109--118, 2019.

\bibitem{toupin2023shaping}
Sophie Toupin.
\newblock Shaping feminist artificial intelligence.
\newblock {\em New Media \& Society}, page 14614448221150776, 2023.

\bibitem{van2022intersectional}
Tom van Nuenen, Jose Such, and Mark Cote.
\newblock Intersectional experiences of unfair treatment caused by automated computational systems.
\newblock {\em Proceedings of the ACM on Human-Computer Interaction}, 6(CSCW2):1--30, 2022.

\bibitem{gender:bias:14}
C.~N. Vorisek, C.~Stellmach, P.~J. Mayer, S.~A.~I. Klopfenstein, D.~M. Bures, A.~Diehl, and S.~Thun.
\newblock Artificial intelligence bias in health care: Web-based survey.
\newblock {\em Journal of Medical Internet Research}, 25(e41089), 2023.

\bibitem{gender:bias:12}
C.~Wang, K.~Wang, A.~Bian, R.~Islam, K.~N. Keya, J.~Foulds, and S.~Pan.
\newblock Do humans prefer debiased ai algorithms? a case study in career recommendation.
\newblock In {\em International Conference on Intelligent User Interfaces}, pages 134--147, 2022.

\bibitem{AI:education:5}
R.~Williams, H.~W. Park, L.~Oh, and C.~Breazeal.
\newblock Popbots: Designing an artificial intelligence curriculum for early childhood education.
\newblock In {\em Proceedings of the AAAI Conference on Artificial Intelligence}, volume~33, pages 9729--9736, 2019.

\bibitem{wright2020ai}
Steven~A Wright.
\newblock Ai in the law: Towards assessing ethical risks.
\newblock In {\em 2020 IEEE International Conference on Big Data (Big Data)}, pages 2160--2169. IEEE, 2020.

\bibitem{gender:bias:8}
J.~Zhao, T.~Wang, M.~Yatskar, V.~Ordonez, and K.~W. Chang.
\newblock Gender bias in coreference resolution: Evaluation and debiasing methods.
\newblock {\em arXiv preprint arXiv:1804.06876}, 2018.

\bibitem{AI:education:4}
A.~Zimmermann-Niefield, M.~Turner, B.~Murphy, S.~K. Kane, and R.~B. Shapiro.
\newblock Youth learning machine learning through building models of athletic moves.
\newblock In {\em Proceedings of the 18th ACM international conference on interaction design and children}, pages 121--132, 2019.

\end{thebibliography}

\clearpage
\appendix
\section{Screening Survey}
\label{screening}

The screening survey asks about correspondents' demographic information and pre-study KQs.

\noindent 1. I am currently a \underline{\hspace{0.5cm}} \\
a. Freshman (first year) in college. \\
b. Sophomore (second year) in college. \\
c. Junior (third year) in college. \\
d. Senior (fourth year) in college. \\
e. Other \underline{\hspace{0.5cm}}

\noindent 2. Your major (fill in NA if it doesn't apply).

\noindent 3. How would you rate your level of knowledge in AI (10-point Likert scale: know nothing to extremely good)?

\noindent 4. How do you gain knowledge in AI (multiple options may apply)? \\
a. Courses. \\
b. Social media. \\
c. News. \\
d. Other \underline{\hspace{0.5cm}}

\textbf{The following are knowledge questions.}

\noindent 5. Choose the scenarios that may contain gender bias (multiple options may apply). \\
a. A girl is repeatedly recommended lipsticks and dolls simply based on her gender. \\
b. An AI system automatically predicts a doctor as male and a nurse as female. \\
c. A healthcare algorithm underestimates women’s needs. \\
d. A chatbot shares discriminatory tweets against women. \\
e. A boy is repeatedly recommended sports gear according to his past purchasing history.

\noindent 6. Choose the AI training phases where gender bias can be introduced (multiple options may apply). \\
a. Data collection. \\
b. Data preprocessing. \\
c. Feature engineering. \\
d. Model training. \\
e. Model usage.

\noindent 7. How gender-equal do you think AI systems are (10-point Likert-scale: Not gender-equal at all to Very gender-equal)? 

\noindent 8. Google search scenario: Google's algorithm auto-fills your search. How much do you think the following word predictions are biased (1: not biased at all, 5: very biased)? \\
a. When you search ``hillary clinton,'' Google auto-fills ``partner; age; young; books.'' \\
b. When you search ``bill clinton,'' Google auto-fills ``net worth; impeachment; presidency; age.''

\noindent 9. Computer vision scenario: A model is trained to realize facial recognition. How biased do you think models trained based on the following data are (1: not biased at all, 5: very biased)? \\
a. The model is trained with 100 boys and 200 girls. \\
b. The model is trained with 201 boys and 199 girls.

\noindent 10. How can one remove potential gender bias from an AI system before deployment (multiple options may apply)? \\
a. Removing biased data points from the training data. \\
b. Adding more data points for underrepresented groups. \\
c. Not using biased characteristics (e.g., gender) as a factor in the training process. \\
d. Using a less biased algorithm for the system. \\ 
e. Including more developers from underrepresented groups in the developing team.

\noindent 11. Please enter your contact information (phone number/email).
\section{Experiment Protocol}
\label{protocol}

\textbf{Introduction}

[REDACTED]

\noindent \textbf{Interview}

\begin{itemize}
    \item Could you tell me a bit about yourself?
    \item Have you come across gender bias in AI systems? (For example, have you been repeatedly recommended boyish or girlish items in movie or shopping websites?) 
    \item Why do you think an AI system might work worse for certain people?
    \item \textit{Google search scenario mentioned in the screening survey.} How do you think gender bias is introduced into this AI system? How do you think gender bias can be mitigated from this AI system?
    \item \textit{Computer vision scenario mentioned in the screening survey.} How do you think gender bias is introduced into this AI system? How do you think gender bias can be mitigated from this AI system?
    \item Are AI gender bias and debiasing methods taught in classes? How? 
    \item Have you created AI systems by yourself? (If so) What kinds? Could you give me some examples? Do you consider debiasing when implementing AI systems on your own?
    \item If you were to deploy an AI system, are you confident that you can deal with potential gender bias in your system? How?
\end{itemize}

\noindent \textbf{Two Tutorials}

Participants are asked to think aloud while completing the tasks in the two tutorials. They are also asked questions about their in-activity behaviors. After each tutorial, participants are asked to fill in the KQ/usability survey. KQs are the same as those in the screening survey. Below are the usability questions (7-point Likert scale: strongly disagree to strongly agree). 
\begin{itemize}
    \item The tutorial was well organized and made good use of time.
    \item I learned about something technical.
    \item I learned about something important for society.
    \item I felt the tutorial was interesting.
    \item I felt the tutorial kept my attention.
    \item The sources of bias introduced were easy to understand.
    \item The debiasing methods introduced were easy to understand.
 \end{itemize}

Several verbal questions are also asked.
\begin{itemize}
    \item How do you feel about this lab? Do you think it is difficult to understand?
    \item Where do you think the gender bias comes from?
    \item How many approaches have you learned so far to mitigate gender bias from AI? Do you think they are easy to understand? Can you think of other ways to mitigate gender bias? 
\end{itemize}

\noindent \textbf{Exit Interview}
\begin{itemize}
    \item How is gender bias introduced in the training and deployment process? Has your understanding changed after our experiment?
    \item How can gender bias be mitigated from AI systems from a developer's perspective? Has your understanding changed after our experiment? Do you think it reasonable to twist social reality to mitigate gender bias?
    \item What's the difference between pre-existing bias (in data) and technical bias (in training)?
    \item What do you think was the most important thing you learned in today's experiments? How could the knowledge you learned today be used in practice?
    \item How do you think about the study/tutorials? What is one thing you'd like us to explain more clearly? How do you like the hands-on nature of our tutorials? Pros and cons? How do you like the scenario-based nature of our tutorials? How do you like the technical nature of our tutorials?

\end{itemize}
% \clearpage
\section{Recruitment Tutorial Screenshots}
\label{tutorial}

\begin{figure}[htbp]
    \centering
    \begin{minipage}{0.75\linewidth}
        \centering
        \includegraphics[width=\linewidth]{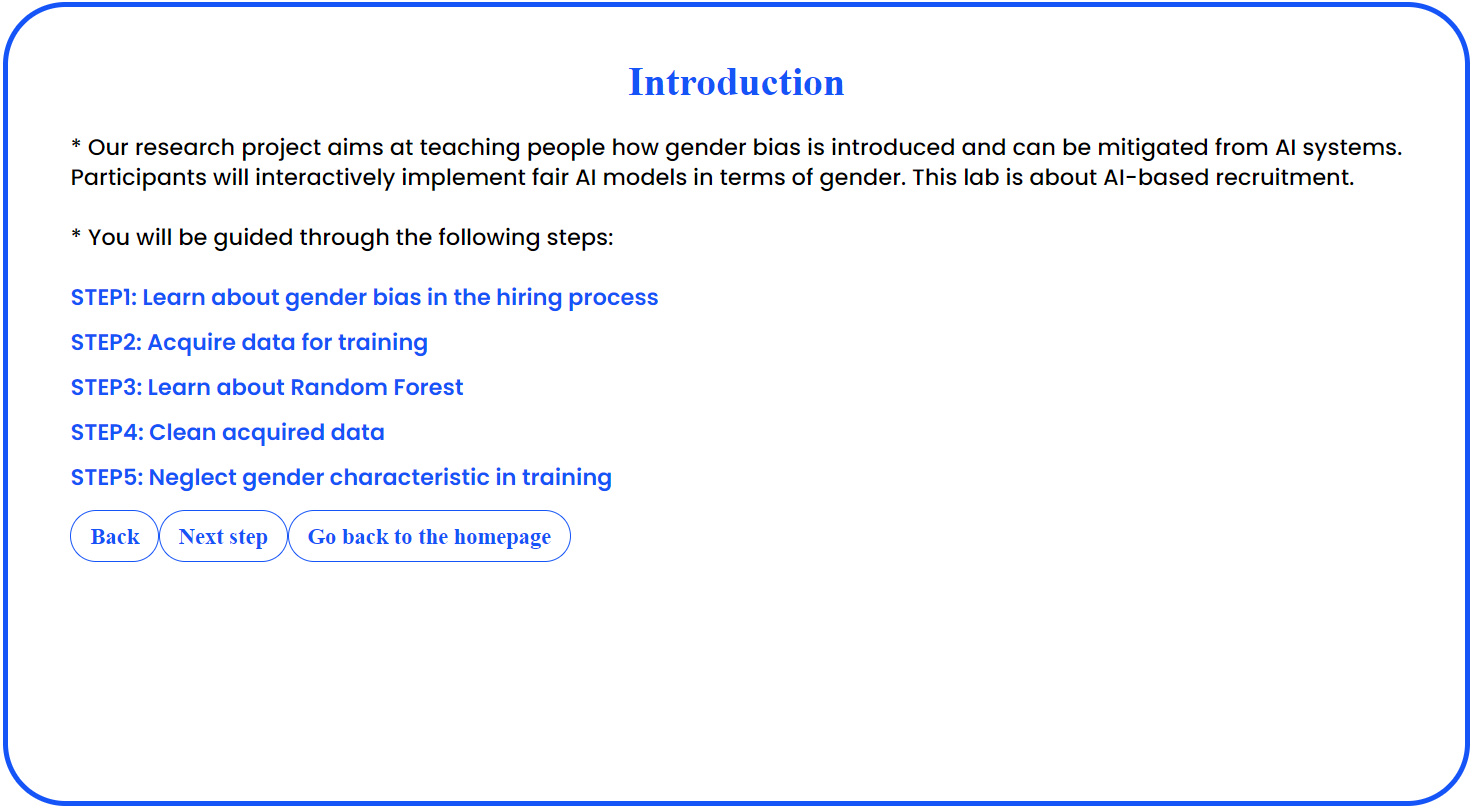}
        \subcaption{Overview.}
        \label{recruitment-overview}
    \end{minipage}%
    
    \begin{minipage}{0.75\linewidth}
        \centering
        \includegraphics[width=\linewidth]{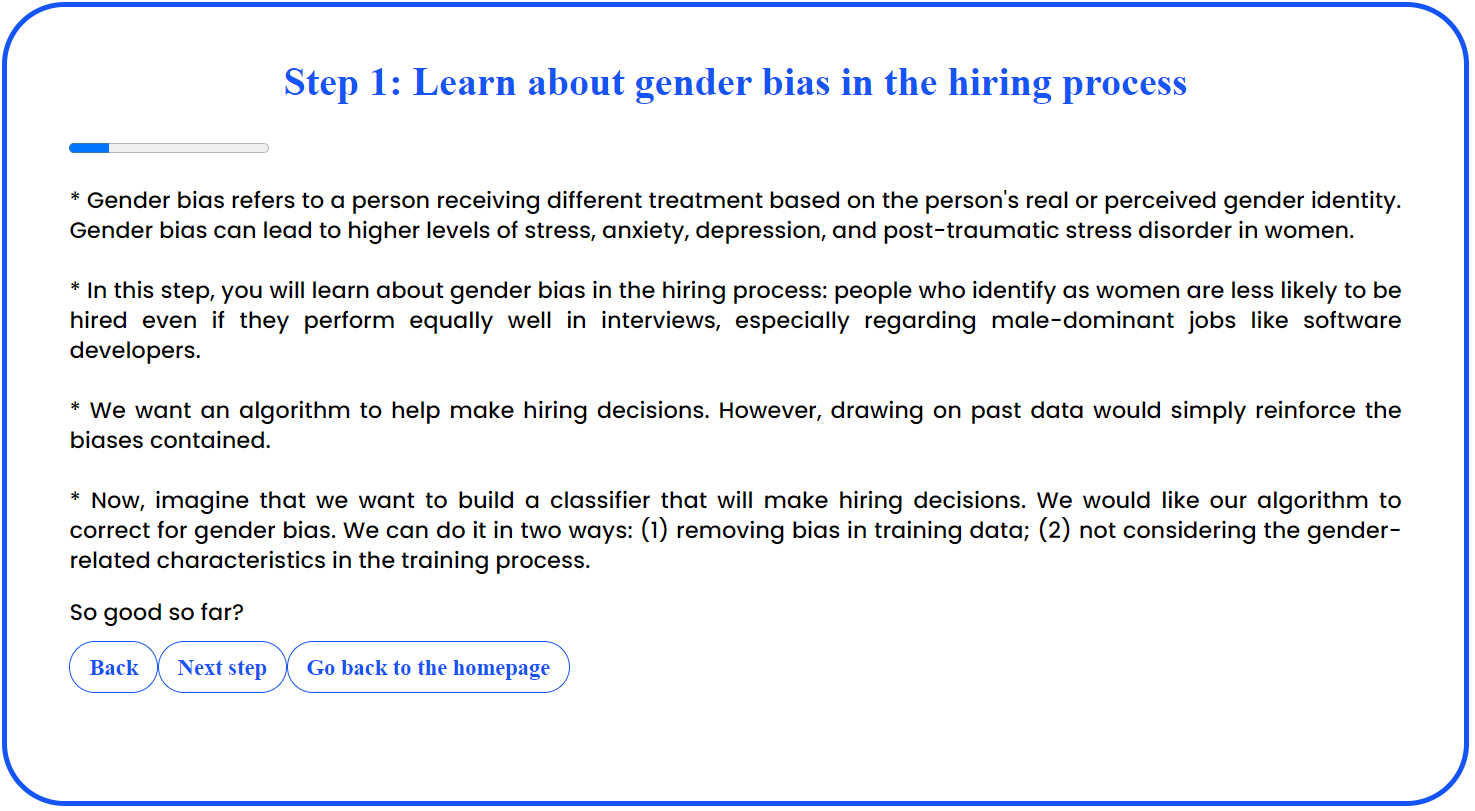}
        \subcaption{Step 1: Learn about gender bias in the hiring process.}
        \label{recruitment-hiring}
    \end{minipage}%
    
    \begin{minipage}{0.75\linewidth}
        \centering
        \includegraphics[width=\linewidth]{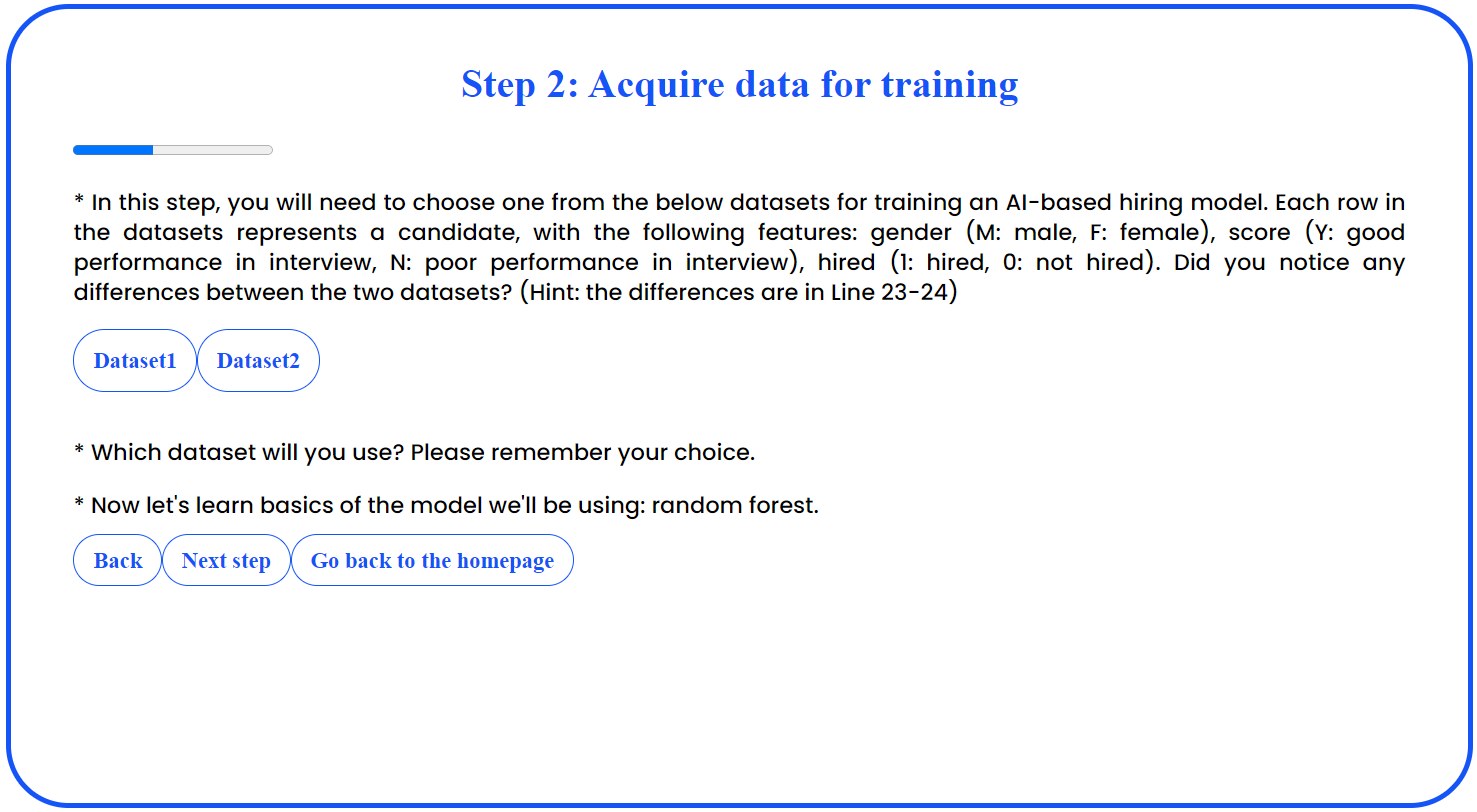}
        \subcaption{Step 2: Acquire data for training.}
        \label{recruitment-dataset}
    \end{minipage}
    \caption{Recruitment tutorial screenshots (Overview - Step 2).}
    \label{tutorial:screenshot:1}
\end{figure}

\clearpage % Force a page break

\begin{figure}[htbp]
    \centering
    \begin{minipage}{0.75\linewidth}
        \centering
        \includegraphics[width=\linewidth]{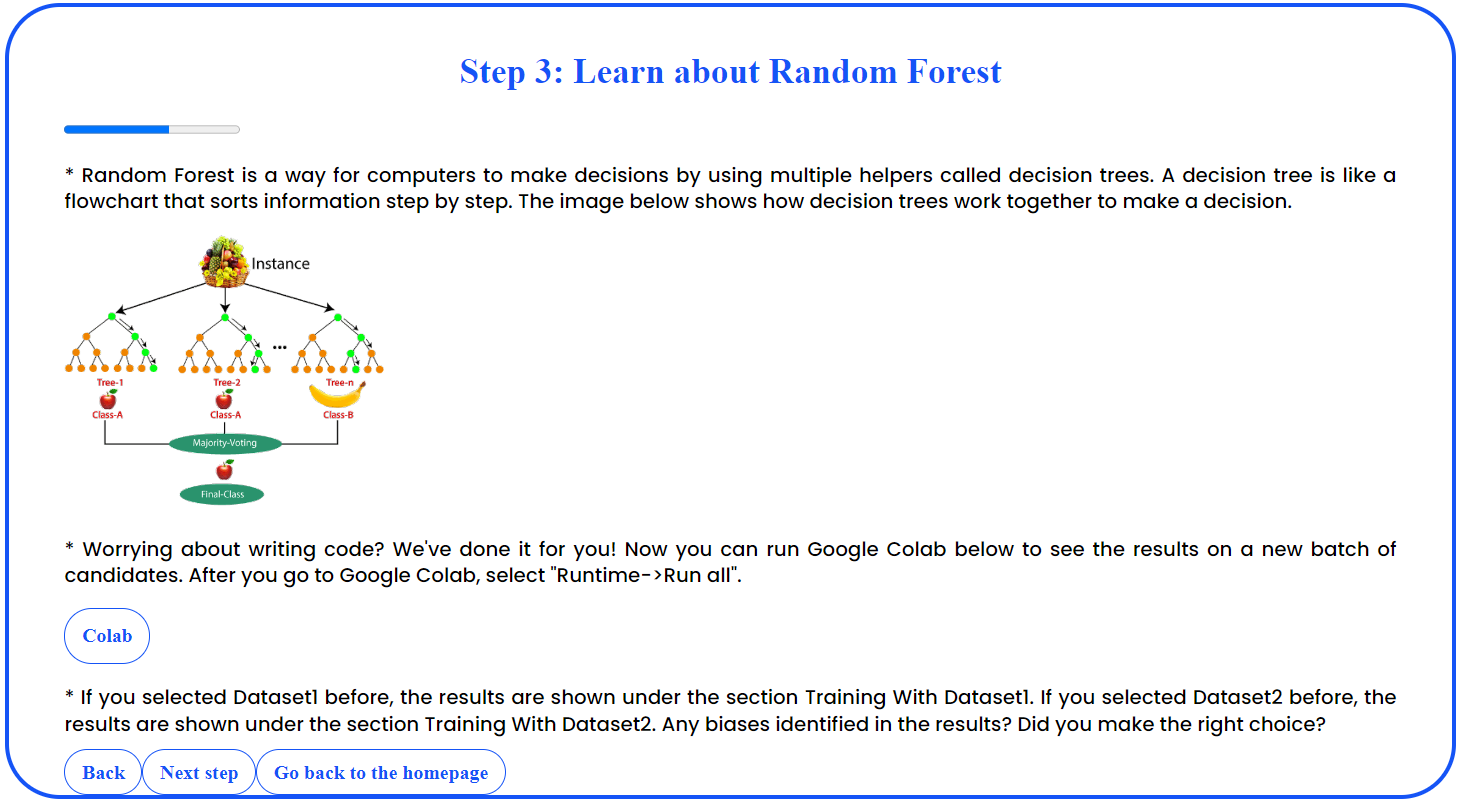}
        \subcaption{Step 3: Learn about Random Forest.}
        \label{recruitment-random-forest}
    \end{minipage}%
    
    \begin{minipage}{0.75\linewidth}
        \centering
        \includegraphics[width=\linewidth]{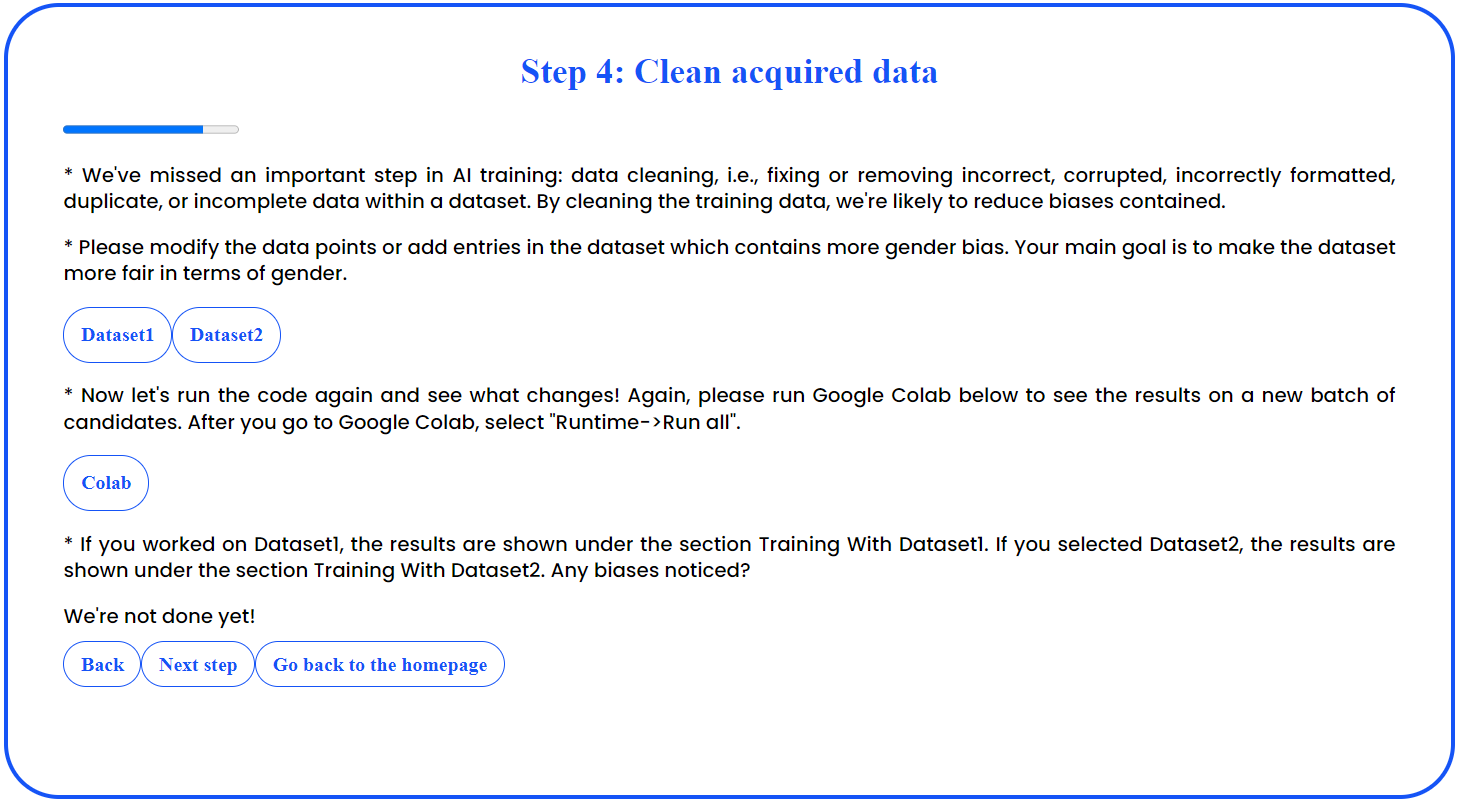}
        \subcaption{Step 4: Clean acquired data.}
        \label{recruitment-debiasing-1}
    \end{minipage}%
    
    \begin{minipage}{0.75\linewidth}
        \centering
        \includegraphics[width=\linewidth]{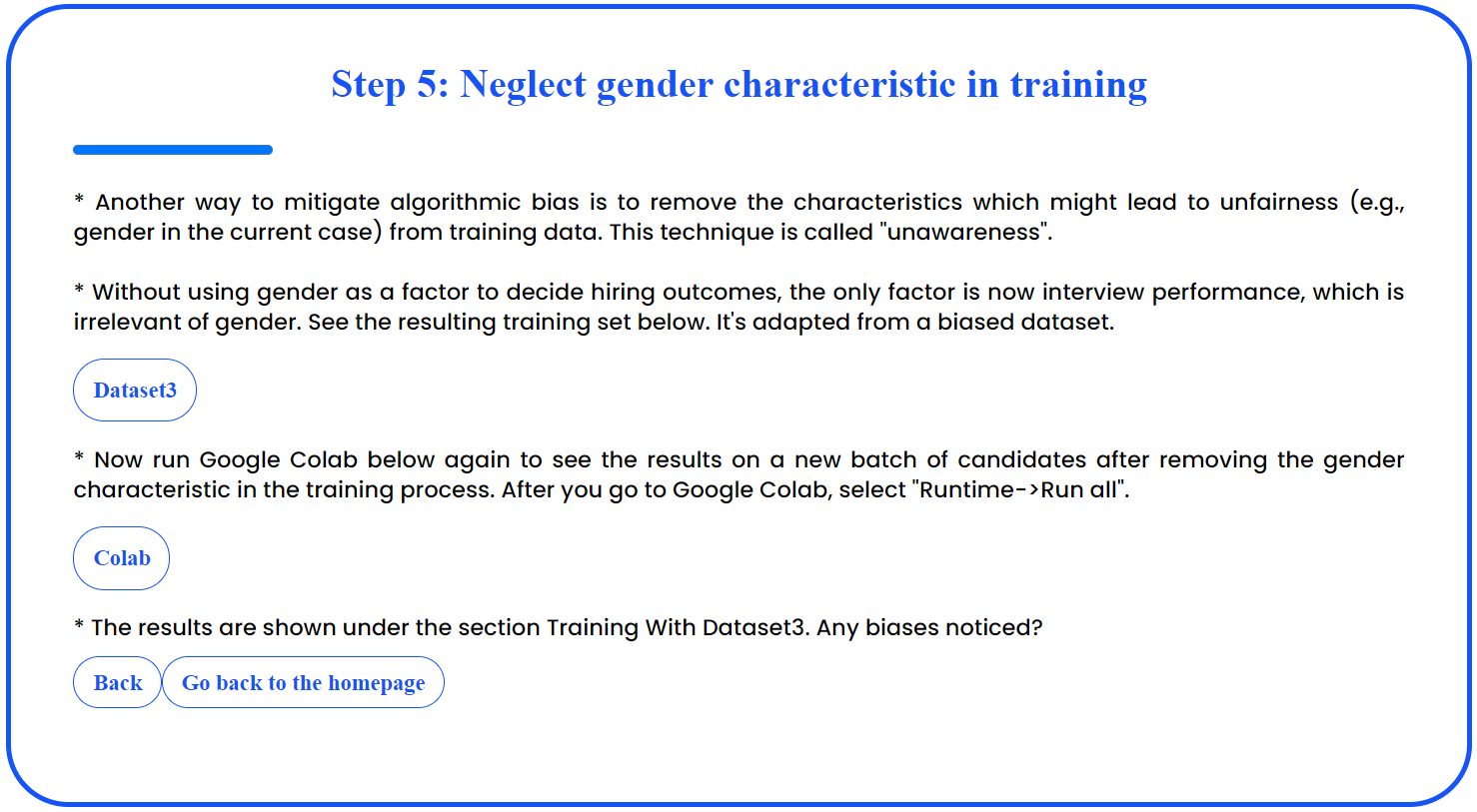}
        \subcaption{Step 5: Neglect gender characteristic in training.}
        \label{recruitment-debiasing-2}
    \end{minipage}
    \caption{Recruitment tutorial screenshots (Step 3 - Step 5).}
    \label{tutorial:screenshot:2}
\end{figure}

\end{document}